# Feasibility of Quantum Key Distribution from High Altitude Platforms

Yi Chu[1], Ross Donaldson[2], Rupesh Kumar[3] and David Grace[1]

[1] Department of Electronic Engineering, University of York, York, YO10 5DD, United Kingdom
[2] SUPA, Institute of Photonics and Quantum Sciences, School of Engineering and Physical Sciences, Heriot-Watt University, Edinburgh, EH14 4AS, United Kingdom
[3] Department of Physics, University of York, York, YO10 5DD, United Kingdom

E-mail: yi.chu@york.ac.uk (corresponding author)

**Abstract**

This paper presents the feasibility study of deploying Quantum Key Distribution (QKD) from High Altitude Platforms (HAPs), as a way of securing future communications applications and services. The paper provides a thorough review of the state of the art HAP technologies and summarises the benefits that HAPs can bring to the QKD services. A detailed link budget analysis is presented in the paper to evaluate the feasibility of delivering QKD from stratospheric HAPs flying at 20 km altitude. The results show a generous link budget under most operating conditions which brings the possibility of using diverged beams, thereby simplifying the Pointing, Acquisition and Tracking (PAT) of the optical system on the HAPs and ground, potentially widening the range of future use cases where QKD could be a viable solution.

## 1. INTRODUCTION

Quantum Key Distribution (QKD) is a potentially revolutionary cryptographic technique which offers theoretically secured cryptographic key delivery between two parties, typically named Alice (the transmitter) and Bob (the receiver). The security of QKD is based on the laws of quantum physics [1]. The key shared by QKD is generated by quantum randomness, rather than an algorithm, meaning the shared key is robust to future advances in decryption algorithms and attacks from quantum computers [1,2]. QKD relies on quantum superposition, quantum uncertainties, and quantum entanglement for secure key distribution/generation as well as identification of eavesdropper activity in the communication channel. These two benefits make QKD an attractive cryptographic technique. Other quantum derivatives of communications protocols also exist, such as quantum digital signatures [3], coin flipping [4], and counterfactual communications [5]. However, QKD is the most mature quantum communications protocol, and is already seeing commercial activity.

QKD over optical fibre links has been an area of active research for decades [6-9]. Optical fibre network demonstrations have also been shown on dark-fibre networks around the globe [10-12]. Due to the exponential loss of optical fibre with distance, long-distance secure key distribution over optical fibre becomes inefficient. Even though the state-of-the-art ultra-low loss fibre is able to achieve 0.14 dB/km loss [13,14], the attenuation is still significant at large distances which results in a limited secure key rate. Multi-hop links based on relay nodes can overcome this limitation [15], however, additional security assumptions are required, for example the relay nodes must be trusted.

The quantum channel in free space communications has a much lower loss over distance. For example, 0.07 dB/km loss has been reached in atmosphere in [16]. QKD via satellites has been considered as an alternative to deliver keys over large distances by utilising the free space quantum channel. Satellites located at less than 2000 km low earth orbit (LEO) provide much less attenuation than fibre at the same distance, thereby achieving higher secure key rate. Experimental demonstrations and feasibility studies have already shown satellite QKD is a viable approach and has the potential of becoming a deployable service [17-21]. However, the high costs of satellite operations and difficulties of equipment maintenance in space will always be the barrier between the technology and the market.

Another method of exploiting the free space quantum channel is QKD via high altitude platforms (HAPs). This approach has not been widely considered because of the immature HAP technology and lack of global deployment capability. The current development of HAPs has a projected 1.5 billion USD market by 2024 [22] and HAPs have been proved to be able to continuously deliver commercial services. Alphabet's Loon [23] has already started providing 4G wireless communications services to remote areas of Kenya [24] by using multiple free floating high altitude balloons.

Free space optics has been used as inter-platform links between balloons. Conventional communication via aerial platforms has been demonstrated feasible by many experiments [25-28] but it is rarely considered as an option for delivering QKD. The work in [29] has demonstrated QKD from a Dornier 228 utility aircraft and the work in [17] has implemented QKD from a hot air balloon but its purpose is to evaluate QKD from satellite as an intermediate step.

Compared with the predictable trajectory of the satellites, the movements of HAPs are more random (because of the wind) which brings more challenges to pointing, acquisition and tracking (PAT) of the optical system. However, the lower link distance provides more tolerance to attenuation and operating potential during daylight which can compensate the disadvantage. The station-keeping and long endurance capabilities of HAPs allow the QKD services to be delivered to certain regions continuously, unlike the unavoidable service window of QKD from LEO satellites. Some HAPs do not even need a specific launching facility (e.g. the Airbus Zephyr maiden flight [30]) which brings the possibility of rapid deployment and removes the regional limits of the service. The lower deployment costs allow the QKD service to be accessible to a larger market. The ease of HAP launch and maintenance can maintain continuous QKD services by using multiple HAPs simultaneously. In this paper, we will review the feasibility of QKD from HAPs to ground, including the challenges and potential solutions of the PAT system and the link budget, and provide a vision of future implementation.

The rest of the paper is organised as follows. Section II reviews the state-of-the-art HAP technologies. Section III gives an overview of QKD technologies. Section IV presents the analysis of link budget under different operating conditions. Section V explains the challenges and potential solutions of PAT system. Section VI concludes the paper.

**2. The HAP Technologies**

Ideally, HAPs are able to continuously cruise in the stratosphere at about 20 km altitude for several months. The renewable energy source equipped by HAPs can harvest energy to power the aircraft and the payload. They can be deployed rapidly and relocate globally according to their applications and tasks. There are two major types of HAPs, heavier-than-air (mainly fixed-wing HAPs) and lighter-than-air (free-floating balloons and airships) aircraft. In this section we will review the state-of-the-art of these HAPs and their properties related to QKD applications.

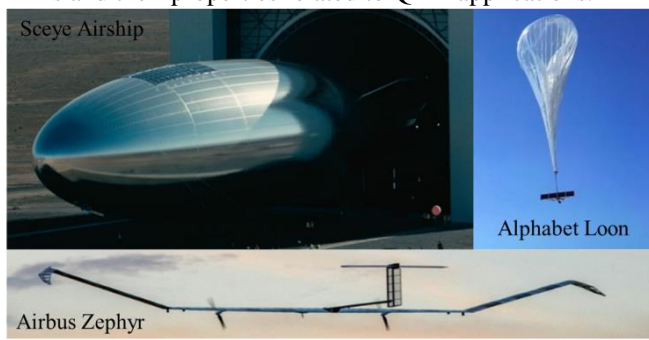

Fig. 1 Sceye airship [75], Alphabet Loon [23] and Airbus Zephyr [30]

*2.1. Fixed-Wing HAPs*

There are many fixed-wing HAPs under development and in operation, they can carry the payload that weights from a few kg to a few hundred of kg. The HAP which can carry the largest payload (680 kg) is the Global Hawk [31] developed by National Aeronautics and Space Administration (NASA), however it is powered by fuel so its duration of a single flight is limited. Unlike the Global Hawk many other fixed-wing HAPs are powered by the renewable solar energy to achieve long endurance. One representative HAP is the Airbus Zephyr S [30], which has kept the record of the longest airborne time of HAPs of almost 26 days. However, given its weight (75 kg) and size (25 m wingspan) its payload capability is limited (2 kg) and the power provided to the payload is up to 200 W and depends on the harvested solar energy.

There are larger HAPs with more payload capacity available and under development. For example, Airbus is developing Zephyr T and Zephyr Future Evolution, which aims to have up to 40 kg payload capacity and 120 days of single mission duration [32]. The PHASA-35 [33] developed by Prismatic/BAE Systems has 15 kg payload capacity and up to 1 kW payload power. It is expected to be airborne continuously for one year without landing at 35 °latitude. The Odysseus [34] developed by Boeing has 25 kg payload capacity, 250 W payload power and several months endurance. The Elektra-2 [35] developed by German Aerospace Centre (DLR) can carry up to 120 kg payload with 5 kW payload power and almost unlimited flight duration. The Stratospheric Platforms [36] is developing a unique HAP powered by liquid hydrogen rather than solar power, which generates over 20 kW of power for the payload [37]. The HAP has 60 m wingspan, 140 kg payload capacity and can fly continuously for 9 days.

In general, the fixed-wing HAPs are powered by solar-electric energy and have long endurance. They are equipped with electric motors which provide lift and thrust so the HAPs can cruise at 20 km stratosphere with the speed of around a hundred km/h. Compared with satellites, the launch and landing of HAPs are relatively flexible (depends on the size of the HAPs) without the requirements of specialised facilities. They can usually reach operating altitude several hours after launching. All these properties facilitate QKD services thereby making this HAP archetype a favourable option:

• Continuity of service: several months of flight duration and rapid deployment ensure that there are no gaps of the QKD service, in principle.

• Ease of maintenance: quick and flexible launch/landing allow the maintenance work to take place without significant costs.

• Relocation and diversity: fixed-wing HAPs can travel to task locations globally based their own power and mobility (latitude dependent), in the event of temporary link outage (e.g. extreme weather completely blocks the optical path) the HAPs can move to locations which are not affected by the weather.

Table II in the Appendix summarises the specifications of the fixed-wing HAPs available in the public domain.

*2.2. Free-Floating Balloons and Airships*

One advantage of lighter-than-air HAPs is that their lift is not originated from the power consuming motors and wings, which makes them much easier to reach and maintain the operating altitude and to achieve the target endurance. Free-floating balloons can be massively deployed benefiting from their low costs, however they can only be relocated according to the different wind directions at different altitudes and their station-keeping is difficult. Alphabet's Loon approaches this problem by deploying multiple balloons to always keep one or more balloons above the service area [23]. Equipment maintenance could also be difficult because of the limited aerial manoeuvrability and the randomness of the landing zone.

The other advantage of lighter-than-air HAPs is that they are easier to scale up to larger size compare with the fixed-wing HAPs, which allows them to carry much heavier payload. For example, the Thales Stratobus [38] weighs 5 tons and carries 250 kg of payload. It has 4 electric motors and can keep stationary while experiencing up to 90 km/h wind. Airships have similar generous payload capacity as the balloons, but much better mobility from the electric motors. The station-keeping, endurance and payload capabilities make the airships another possible carrier of the QKD service. However, the larger airships are difficult to maintain and repair, and they require specialised ground facilities for launching, landing and storage. There are also some low-cost low altitude platforms developed for certain applications and evaluation purposes. For example, the British network operator EE has demonstrated 4G service delivered from a Helikite [39] to an event in rural Wales [40]. Table III in the Appendices section lists the specifications of some lighter-than-air aerial platforms available in the public domain.

**3. Quantum Key Distribution**

In this section, we provide a general description of QKD, where the protocol for key distribution is separated into two sections: quantum communication for quantum signal generation, transmission and detection; and classical communication for post processing the data from the quantum communication session.

In quantum communication session Alice generates a long sequence of random quantum signals, either from a set of quantum states with pre-defined classical bit values or from a distribution with undefined classical bit values. The former is the case of Discrete-Variable (DV) QKD, where detection of quantum states reveals the encoded key bits. While in the latter case referred as Continuous-Variable (CV) QKD, direct measurement of the quantum signal does not reveal the key, but the data-post processing establishes a common key between Alice and Bob. In the following we will explain the quantum communication sessions of DV-QKD and CV-QKD and then describe the classical communication which is more or less common to both QKD systems.

*3.1. Quantum Communication with Discrete Variable*

In DV-QKD, the key information is encoded on discrete degrees-of-freedom of the quantum optical states [41][42]. The quantum states can be generated using pseudo-deterministic [43] or probabilistic entangled photon sources [44]. Quantum state superposition or quantum entanglement are utilised for the secure transmission of the key and reveal eavesdroppers in the communications channel. In optical-fibre implementations, protocols based on phase [10][45] and time-bin [10] encoding are traditionally used, as those degrees-of-freedom are more robust to transmission in optical fibre. Polarisation based protocols, such as the BB84 protocol [46], are traditionally used in free space communications, due to the robustness of polarisation to atmospheric transmission [47].

As an example of a protocol in operation, the BB84 protocol is used here, Alice has two sets of paired attenuated laser sources, which are used to encode the four quantum states. Individual lasers within a pair are used to directly encode the binary key bits. Having two sets of paired lasers enables two basis sets for encoding and decoding, creating the quantum superposition states. Alice uses a quantum random number generator (QRNG) to select one basis set for encoding, and uses the QRNG again to select the random key bit. Alice records her basis set choice and the key bit, and transmits the encoded quantum state to Bob through an optical channel.

When Bob receives the quantum state from the optical channel, he has no *a-priori* information about the quantum state encoded by Alice, and uses his own QRNG to select the decoding basis set. He then records the measurement outcome, using single-photon detectors, as well as his own basis set selection. That information is stored for further processing of the key. In this paper, we present a link budget analysis of a polarisation based weak coherent pulse decoy state BB84 protocol [46].

*3.2. Quantum Communication with Continuous Variable*

In this description of continuous variable communication for QKD, we consider Gaussian modulated coherent state protocol (GMCS) [48][49] in which Alice generates sequence of random amplitude and phase modulated coherent state $|\alpha> = re^{-i\theta}$ such that the distribution of the quadrature, $X_A = r\cos\theta$, and $P_A = r\sin\theta$, follows normal distribution with variance, $V_A$, and mean zero. Here, $|r|^2$ is the intensity of the coherent signal which corresponds to a few photons per pulse, on average. And $\theta$ is the relative phase of the coherent signal with respect to an intense reference signal referred as local oscillator (LO). It is either generated at Alice and sent to Bob along with the coherent state referred as transmitted local oscillator (TLO) scheme, or generate locally at Bob referred as local local oscillator (LLO) scheme. In LLO scheme, since it uses two different lasers, one at Alice and another at Bob, it requires the establishment of a common phase reference between the users which is achieved by sending a phase reference pulse, $R_{ref}$ from Alice to Bob.

Bob randomly measures one of the quadrature components using a shot noise limited homodyne receiver. This is performed by mixing the input quantum signals with an intense LO on a symmetric beam splitter. The output of the beam splitter is individually detected using reverse biased PIN photodiodes, the photocurrents are then subtracted from each other and amplified. The amplified output represents a noisy version of Alice's quadrature values $X_A$ or $P_A$, depending on 0 or 90$^o$ relative phase with respect to LO. This would create a correlated data set of quadrature value between Alice and Bob for the raw key.

## 3.3. Classical Communication for Post-processing

Once Bob has registered the measurement outcome which is the raw key of the quantum communication session, the users have to sift the key in order to match the basis of the quantum signal generation and detection at both ends. Once the sifting has been done, Alice shares a part of her sifted key with Bob which he uses to compare with the sifted key in his possession. This reveals the quantum bit error rate (QBER) of the DV-QKD protocol, a signature of eavesdropping.

In CV-QKD, instead of QBER, comparing the variance of a part of the sifted quadrature values reveals the presence of eavesdropping as noise which is called excess noise variance. Unlike DV-QKD, since CV-QKD quadrature values are analogue values, and additional post-processing is employed to convert the analogue values to binary digits. If the QBER or excess noise are below the permissible limit for secure key generation, Alice and Bob apply classical error correction techniques such as: cascade or low-density parity check on the rest of the sifted key and convert it to an error corrected key, which is ideally a perfectly correlated string of bits. Finally, to reduce the information leaked to an eavesdropper during the quantum communication session as well as from classical post processing, they apply universal hashing on the error corrected key in order to amplify the privacy of the final key. The amount of privacy amplification, reduction in the size of the error correcting key, is decided on the estimation of eavesdropped information from QBER or excess noise.

In generic form, the final key rate equation can be written as $K = I(A:B) - \min\{I(A:E), I(B:E)\}$. Here, $I(A:B)$ is the mutual information between Alice and Bob, $I(A:E)$ is the information between Alice and Eve (the Eavesdropper), which has to be taken into account for direct reconciliation where Bob correct his noisy measurement outcomes with respect to Alice. $I(B:E)$ is the information between Bob and Eve, in the case of reverse reconciliation where Alice correct her data in order to match with the noisy version of Bob's measurement outcomes.

Giving a detailed description of estimating the mutual information and eavesdropped information is beyond the scope of this paper, so the description is restricted here to the general perspective. Please refer to [50] and reference therein for detailed QKD theoretical analysis.

## 4. Link Budget Analysis

As with conventional communication systems, QKD requires the link budget to be closed to have enough photons arriving at the receiver telescope to transmit the keys. There are many factors affecting the link budget, including the transmission distance, wavelength, optical design, time of the day, optical components, weather, channel turbulence, and background noise. In order to achieve continuity of service, it is important to have the link budget closed with these varying factors. This section will provide details of the variables affecting the link budget and analyses the major operating conditions. The findings from this section will lead to an estimate of QKD performance.

## 4.1. Field of View (FoV) and Background Noise

QKD, and indeed all quantum communication protocols, have a performance dependence on link budget and background noise. If the parameters of the optical components at the receiver are known, the power of the background noise $P_b$ varies according to the brightness of the day [51]:

$$P_b = H_b \times \Omega_{fov} \times A_{rec} \times B \quad (1)$$

where $H_b$ is the brightness of the day, $\Omega_{fov}$ is the receiver FoV, $A_{rec}$ is the area of the telescope aperture and $B$ is the bandwidth of the optical filter. $H_b$ varies at different time of the day, and the typical values are 150 (daytime with illuminating cloud), 15 (hazy daytime), 1.5 (clear daytime), $1.5 \times 10^{-3}$ (full moon night), $1.5 \times 10^{-4}$ (new moon night) and $1.5 \times 10^{-5}$ (moonless night) Wm$^{-2}$Sr µm [51]. Additional background from light pollution can be added if the level of light pollution at a receiver's location is known. Across the range, there is maximum 70 dB difference in the background noise power, which highlights one of the major challenges in operating QKD during daytime.

The state-of-the-art optical filters can achieve 0.1 nm or better bandwidth, and the typical value 0.1 nm will be used in the link budget computation later in this section. Quantum optical states can be generated with a narrow bandwidth, justifying the filter choice. With temperature stability on the HAP, the wavelength of the quantum optical states can be kept within the window of the optical filter. The velocity of the HAP platforms is also not large enough to cause a significant Doppler shift in wavelength.

The FoV of the receiver determines the amount of light (noise and the desired signal) collected by the telescope that reaches the detector. The optical receiver is normally a multi-lens system so obtaining the accurate FoV could be difficult without the detailed design of the system. A common receiver design uses a Schmidt-Cassegrain telescope followed by a collimation lens to produce a collimated beam for the downstream optical components. We can make the assumption that changing the receiver telescope aperture (with the same focal ratio) does not affect the rest of the system. The receiver can be considered as a two-lens system where the first lens is the telescope and the second lens represents the rest of the optical components (which remain the same). The FoV of a lens can be expressed as [52]:

$$\Omega_{fov} = 2 \times \tan^{-1}\left(\frac{D}{2F}\right) \quad (2)$$

where $D$ is the lens diameter and $F$ is the focal length. In our case $D$ is the detector diameter and $F$ is the effective focal length of the optical system. The effective focal length of a two-lens system can be obtained by [52]:

$$F = \frac{f_1 \times f_2}{f_1 + f_2 - d} \quad (3)$$

where $f_1$ is the focal length of the telescope, $f_2$ is the focal length of the other lenses and $d$ is the distance between two lenses. Based on the previous assumptions, when varying the telescope aperture size, $f_1$ changes linearly with the aperture size, the terms $f_2$ and $f_1 - d$ remains the same. The effective focal length of the system changes linearly with the aperture size. We can then conclude that the FoV $\Omega_{fov}$ decreases linearly with an increasing aperture size, which captures of benefits using larger telescopes (receiving less background noise directly on the quantum detector(s)).

## 4.2. Channel Loss

There is major channel loss (geometric loss) resulting from the natural spreading of the beam [53]. The geometric losses are typically the dominant losses in a free space QKD implementation. It can be expressed as:

$$L_{geo} = 20 \log_{10}\left(\frac{D_{tx}+R_{LoS}\times\theta}{D_{rx}}\right) \quad (4)$$

where $D_{rx}$ and $D_{tx}$ are the receiver/transmitter telescope aperture size, $R_{LoS}$ is the line-of-sight (LoS) distance between the two optical terminals, $\theta$ is the beam divergence. In the HAP scenario $R_{LoS}$ can be computed by:

$$R_{LoS} = \frac{H_{HAP}}{\sin\alpha} \quad (5)$$

where $H_{HAP}$ is the altitude of the HAP and $\alpha$ is the elevation angle which varies between 0° and 90°. $\theta$ can be computed by [54]:

$$\theta = 1.22\frac{\lambda}{D_{tx}} \quad (6)$$

where $\lambda$ is the operating wavelength.

The optical link between the ground and the HAP propagates in the atmosphere, which will experience molecular absorption $L_{ma}$ caused by the molecules of water and carbon dioxide [55]. The amount of attenuation depends on the link distance and wavelength, some typical values of $L_{ma}$ are provided in [53]: 0.13 dB/km at 550 nm, 0.01 dB/km at 690 nm, 0.41 dB/km at 850 nm and 0.01 dB/km at 1550 nm.

Different weather conditions cause attenuation when the optical signal propagates through the atmosphere. Fog causes significant attenuation because its particle size is comparable to the wavelength of the optical source. Large snowflakes can potentially block the optical path completely. Visibility range dependent empirical models of attenuation caused by fog, rain and snow are provided in [53]:

$$L_{fog} = \frac{3.91}{V}\left(\frac{\lambda}{550}\right)^{-p} \text{(dB/km)} \quad (7)$$

where $V$ is the visibility range in km, $p$ is the size distribution coefficient of scattering given by:

$$p = \begin{cases} 1.6 & V > 50 \\ 1.3 & 6 < V < 50 \\ 0.585V^{\frac{1}{3}} & V < 6 \end{cases} \quad (8)$$

The attenuation of snow is given by:

$$L_{snow} = \frac{58}{V} \text{ (dB/km)} \quad (9)$$

The attenuation of rain is given by:

$$L_{rain} = \frac{2.8}{V} \text{ (dB/km)} \quad (10)$$

Fig. 2 shows the resulting attenuation under different weather conditions against the visibility range. It can be observed that once the visibility range is falls below 2 km, the rain and snow attenuation increases significantly. Snow always causes large attenuation due to the size of the snowflakes which prevents the implementation of the QKD link. In the HAP scenario we should consider the distance that the optical signal propagates in weather $R_w$, which can be computed by:

$$R_w = \frac{H_w}{\sin\alpha} \quad (11)$$

where $H_w$ is the altitude of the weather, which varies at a few km with rain and snow, or sub-km with fog.

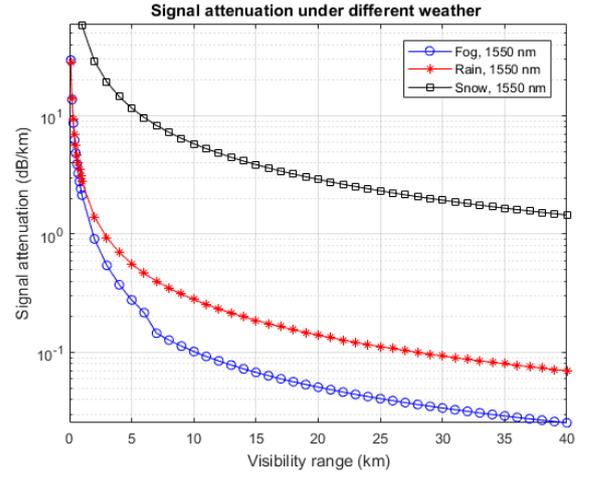

Fig. 2 Optical signal attenuation under different weather conditions

## 4.3. Other Types of Attenuation

There are different types of attenuation resulting from inside the optical system. The performance of the PAT system may affect the link budget significantly when the beams are narrow. The random movements and vibrations of HAPs could potentially cause difficulties for the PAT system to achieve accurate alignment of narrow beams. The attenuation due to misalignment is given as [53]:

$$L_p = exp\left(\frac{-8\theta_j^2}{\theta^2}\right) \quad (12)$$

where $\theta_j$ is the divergence angle of the pointing jitter. The other effect which could cause similar misalignment error is the beam wander. When propagating through the turbulent atmosphere, the beam experiences random deflection caused by the turbulent eddies and the centroid of the beam is randomly displaced [53]. The displacement variance (in m$^2$) can be computed as:

$$L_p = 0.54R_{LoS}^2\left(\frac{\lambda}{D_{tx}}\right)^2\left(\frac{D_{tx}}{r_0}\right)^{\frac{5}{3}} \quad (13)$$

where $r_0$ is the atmosphere field parameter.

Optical components at the receiver can also bring additional attenuation $L_{rx}$, [29]. In this paper, we only consider losses of non-ideal optics 3 dB, and telescope loss, 2.2 dB. These losses are used in the link budget analysis as a benchmark for the DV-QKD protocol.

## 4.4. Link Budget

Summing the different attenuations in previous subsections we can obtain the total loss as (method 1):

$$L_T = L_p + L_{geo} + L_{ma}R_{LoS} + L_wR_w + L_{rx} \quad (14)$$

where $L_w$ is the conditional attenuation caused by different weather (all losses in dB).

Another method from NanoBob [59] estimates the loss as:

$$L_{Nano} = 10\log_{10}\left(\frac{R_{LoS}^2(\theta^2+\theta_{atm}^2)}{D_{rx}^2+T_tT_pT_r}\right)+L_{atm}+L_wR_w+L_{rx} \quad (15)$$

where the beam divergence $\theta$ is estimated twice as in (6):

$$\theta = 2.44\frac{\lambda}{D_{tx}} \quad (16)$$

and $\theta_{atm}$ is the atmosphere turbulence included divergence angle computed as:

$$\theta_{atm} = 2.1\frac{\lambda}{r_0} \quad (17)$$

The term $L_{atm}$ is the atmospheric attenuation due to Rayleigh scattering and absorption (3 dB is given as a typical value), the three terms $T_t$, $T_r$ and $T_p$ are the efficiency of the transmitter telescope, receiver telescope and pointing (all are given 0.8 as typical values).

Table I summarises the parameters used in the link budget analysis, and the parameters apply to the rest of the paper unless specifically mentioned. Note that the channel loss results presented in later sections of the paper have excluded the detector loss $L_{rx}$ from equations (14) and (15), because it has already been incorporated while calculating QBER.

TABLE I
PARAMETERS OF LINK BUDGET ANALYSIS

| Parameter | Value |
| --- | --- |
| Wavelength | 1550 nm |
| Field parameter $r_0$ | 0.2 m |
| Transmitter telescope aperture size | 0.1 m |
| Receiver telescope aperture size | 0.4 m |
| Transmitting power (non-QKD) | 1 mW |
| Divergence of pointing jitter $\theta_j$ | 5 μrad |
| HAP altitude | 20 km |
| HAP elevation angle | 5 ° to 89 ° |
| Fog altitude | 500 m |
| Rain/snow altitude | 5 km |

The HAP altitude and elevation angles together cause the LoS link distance to vary from 20 km to 230 km. Fig. 3 shows the channel loss $L_T$ and $L_{Nano}$ (excluding $L_{rx}$) of both link budget methods at different LoS link distances (weather conditions not included). The NanoBob method has slightly higher loss across most link distance range, partially resulted from the overestimated beam divergence (see equation 16). At the regular HAP operating elevation angles (20 ° or higher, equivalent to 60 km or less LoS distance), the channel loss is 12 dB or less.

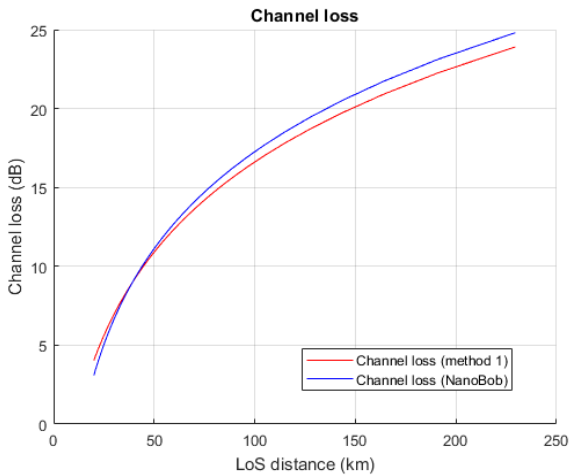

Fig. 3 Channel loss at different LoS distances (weather conditions not included)

Fig. 4 shows the channel loss $L_T$ and $L_{Nano}$ (excluding $L_{rx}$) with different levels of fog (500 m above the ground) existing near the ground receiver. Fig. 5 shows the channel loss with different levels of rain (5 km above the ground). The overall trend of the channel loss is similar to the situation with fog.

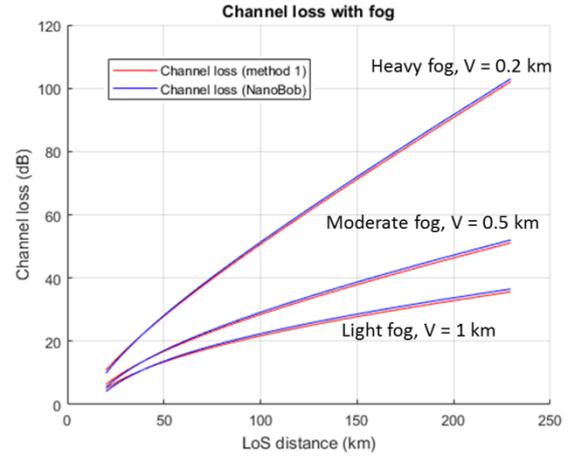

Fig. 4 Channel loss with the presence of fog

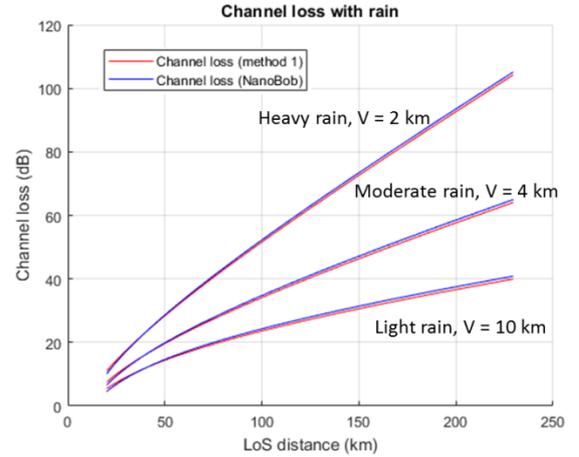

Fig. 5 Channel loss with the presence of rain

### 4.5. Feasibility of QKD

In order to evaluate the feasibility of DV-QKD from HAPs, the decoy-state BB84 protocol was chosen to operate at a moderate frequency of 500 MHz. It was a symmetric basis state protocol, with a quantum signal and one decoy signal, the mean photon numbers were 0.5 and 1, with probabilities 0.8 and 0.2 respectively. The receiver's detectors were chosen to be high performance InGaAs single-photon avalanche diodes (SPADs). The SPADs had a single-photon detection efficiency of 25%, a detector dead time of 18 μs, a detector size of 64.5 μm (fibre core diameter coupled to detector), and a dark count rate of 500 counts per second [56].

The QBER calculation was simplified to only consider the background count rate, as other contributions could not be experimentally defined for this paper. To reduce the background noise, a time-gate filter of 500 ps was applied. At the operational frequency of 500 MHz, this meant that the background noise could be reduced to 25% the expected value. No losses occurred from this time-filtering, as it was expected that modern SPADs would have time-responses narrower than 500 ps [58].

Fig. 6 shows the QBER at the different time of the day with varying background noise levels (varying brightness of the sky). The figure also highlights the maximum QBER bound for the decoy state BB84 protocol, 11%. Beyond that bound, no secure key can be shared. Attenuation due to weather conditions is not considered in either figure. For the night time scenarios (note that the QBER are similar across the three night time scenarios), DV-QKD is able to operate with up to 52 dB channel loss (the loss of the detector is already incorporated). For the three day-time scenarios, DV-QKD is able to operate with up to 43 dB, 34 dB and 24 dB channel loss respectively. Together with the channel loss results in Fig. 3, it can be concluded that the system is robust to other sources of attenuation when operating under all scenarios. The night-time scenarios would be ideal for implementing QKD protocols, primarily because of the reduced background noise.

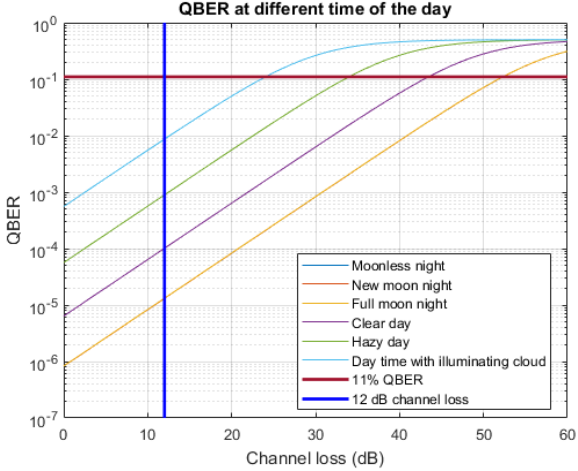

Fig. 6 QBER at different times of the day (weather conditions not included)

Considering the results of Fig. 4 and Fig. 6, the DV-QKD system is able to operate with any levels of fog at almost any time within the regular HAP operating elevation angles. For the day-time with illuminating cloud scenario, the system can operate with the presence of moderate or light fog, but the range is reduced with the presence of heavy fog. Similarly, the DV-QKD system is able to operate with any level of rain at almost any time within the regular HAP operating elevation angles.

## 5. Challenges of PAT on HAPs

In this section we discuss the challenges and difficulties of implementing the PAT system on HAPs. According to the current state-of-the-art HAP technologies, weight and dimension of the payload has strict requirements on fixed-wing HAPs and the situation is less intense on lighter-than-air HAPs. When applying QKD from satellites, beacons are widely used in the PAT systems to ensure that the narrow beams can be accurately aligned with the OGS telescope. In this case equipment such as InGaAs cameras, quadrant detectors and high precision gimbals are used at the transmitter to ensure the accuracy of the PAT. The operating conditions are different when applying QKD from HAPs. Although the HAPs operate at the stratosphere with relatively low wind speed, yet they suffer more random movements and vibrations than satellites. With the much shorter link distances these random movements generate high angular speed and acceleration to the beam, which are more difficult for the PAT system to correct. Alternative methods are desired to lower the difficulty in design and operation of the PAT system while meeting the weight and dimension requirements of the HAPs.

From the link budget analysis in the previous section, system robustness can be observed in most operating scenarios. This indicates that we can trade additional attenuation for the higher tolerance of the PAT system by diverging the beam. When the beam is able to create large footprint at the OGS, beacons may no longer be needed because the coarse PAT using Differential Global Positioning System (DGPS) provides sufficient precision. The high-precision DGPS is able to provide centimetre-level accuracy [60] of the positions of the HAP and the OGS, which are sufficient for diverged beams. Moreover, removing the beacons (and the associated components) will reduce the HAP payload weight significantly.

To diverge the beam we can either use a TX telescope with small aperture size or apply a diverging lens to the beam. Fig. 7 shows the channel loss with varying TX/RX telescope aperture size when the HAP operates at 20 ° elevation. As stated in equation (6) increasing the TX telescope aperture size reduces the beam divergence. We should expect reduced channel loss while using a larger TX telescope however the results of method 1 are showing the opposite when using TX telescopes larger than 0.12 m while the NanoBob method shows the expected performance. The difference is caused by the ways that both methods capture the pointing errors. Method 1 uses equation (12) to estimate the attenuation due to misalignment and the NanoBob uses the fixed pointing efficiency $T_p = 0.8$. The divergence of the pointing jitter $\theta_j$ is 5 μrad (provided in Table I) and this is related to the precision of the PAT system (e.g. the gimbal) which is not dependent on the size of the telescopes. When decreasing the beam divergence $\theta$ the attenuation $L_p$ increases and makes $L_p$ a dominant factor in equation (14) thereby increasing the channel loss. These results also indicate the trade-off between high-cost high-precision PAT system with narrow beams and low-cost high-tolerance PAT system with wide beams. The channel loss of NanoBob method is less affected by the different RX aperture size because in equation (15), the dominant factor of the denominator inside the logarithm is $T_t T_p T_r$ rather than $D_{rx}$ in equation (4).

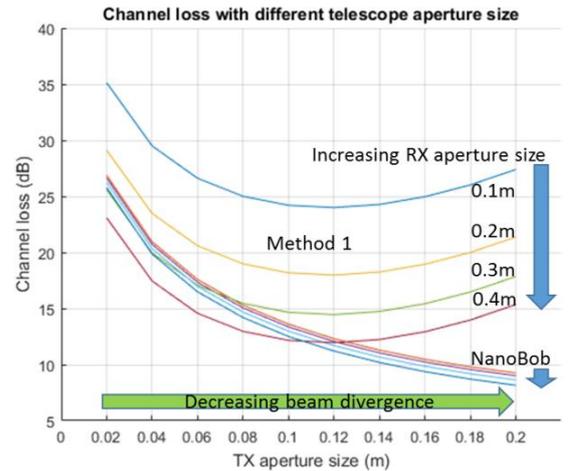

Fig. 7 Channel loss with varying telescope aperture size

Fig. 8 presents the channel loss with different beam divergences during a moonless night with 1 mW transmitted signal power using a 0.1 m transmitter telescope and a 0.4 m receiver telescope. Together with the QBER model presented in Fig. 6, the QBER with different beam divergence can be obtained (shown in Fig. 9). It can be observed that within the regular operating elevation angle of the HAP (equivalent to 60 km or less LoS distance) DV-QKD can remain operational with up to almost 3 mrad beam divergence. The QBER of the 5 mrad case and the 10 mrad case indicates that the current system may not operate correctly and adjustments are required, for instance a larger aperture receiver telescope, narrower optical filtering, a reduction in the dark count rate of the detector, or an increase in the operational frequency of the protocol. For CV-QKD, we consider the Gaussian modulated coherent state protocol with homodyne detection. In order to evaluate the feasibility of CV-QKD, we estimate the signal to noise ratio (SNR) at different channel loss. The signal strength at the output of Alice is set to $10N_0$, system excess noise to $0.03N_0$ and electronic noise variance to $0.1N_0$. Here $N_0$ is the shot-noise variance. Fig. 10 shows the feasibility of CV-QKD at various link distances with different beam divergence values. The threshold for generating positive key rate is limited to SNR of 0.024 below which it is not possible to extract secure keys. We have not considered the effect of background noise effect on the CV-QKD, in this paper.

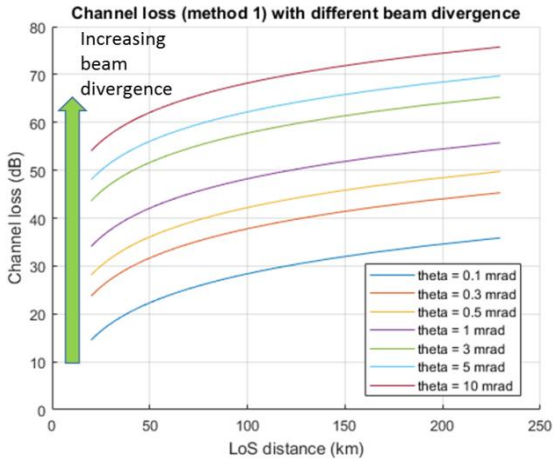
Fig. 8 Channel loss with varying beam divergence

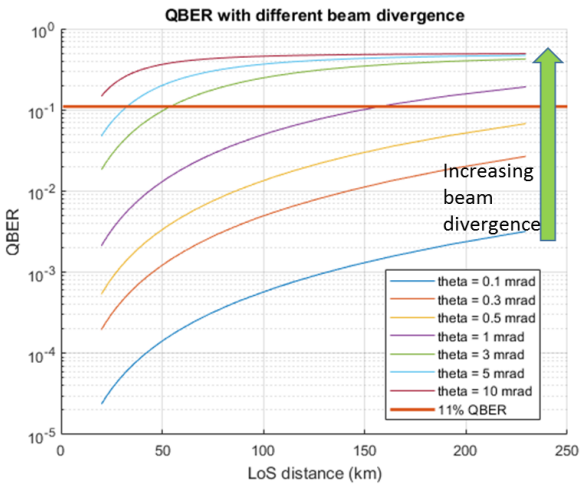
Fig. 9 QBER with varying beam divergence for DV-QKD

The beam with larger beam divergence also provides the opportunity of using low-cost gimbals in the PAT system. Many low-cost off-the-shelf gimbals have the pointing precisions on the level of 0.1 mrad so using mrad level beam divergence can minimise the beam misalignment caused by the low pointing precisions. The difference between the precision of the DGPS signal and the size of the ground beam footprint also contributes to the tolerance of the overall pointing accuracy. For example in Fig. 11 the 1 mrad beam and 3 mrad beam result 10 m and 30 m radius ground beams

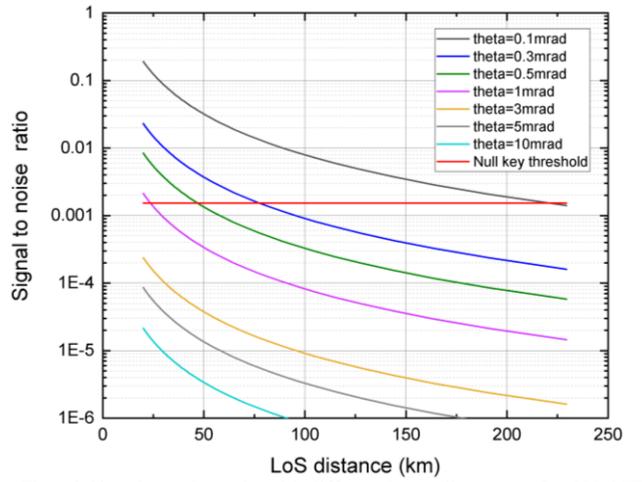
Fig. 10 Signal to noise ratio with different beam divergence for CV-QKD

respectively, which are all magnitudes larger than the centimetre precision of DGPS. These indicate that the pointing precisions of the low-cost gimbals and the precision of the DGPS signals can all be tolerated when using a diverged light source.

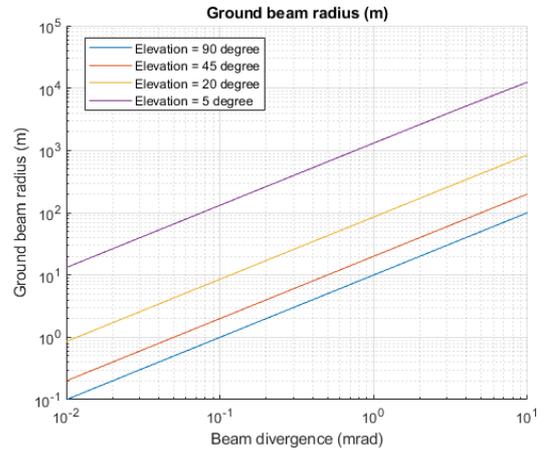
Fig. 11 Ground footprint size against beam divergence

## 6. Conclusion

This paper has presented the feasibility study of delivering QKD from stratospheric HAPs exploiting diverged beams when compared to delivering QKD from satellites. This potentially widens the range of use cases where QKD can operate to secure communications applications and services, while also complementing satellite delivery. HAPs and HAP technologies have been developing rapidly in recent years, which means that both QKD and HAP technologies will be ready for commercial exploitation at similar times. Diverged beams are possible due to HAPs being closer to the ground than satellites, meaning that the optical signal from the HAP naturally suffers less attenuation when compared with satellites of the same specification, thereby having better reliability against situational losses (e.g. different weather conditions). This paper has presented detailed link budget analysis under different operating conditions and the results have shown closed link budget in almost all cases. Compared with satellites, the HAPs have less stability and less predictable trajectories, which requires accurate and frequently updated PAT system to correctly point the beam, if the same level of divergence is used as with the satellite. This paper has proposed a potential method using diverged beams to up to 3 mrad divergence, thereby lowering the requirements of the PAT system and trading additional signal attenuation with greater tolerance of the PAT. As the HAP platform is a more permanent platform, it is thought that this trade-off will have minimal effect on overall service. Simulation results have shown that the link budget can still be closed while using larger beam divergence. This indicates that PAT systems with lower specifications can be used on HAPs, thereby making the payload easier to fit in the weight and dimension requirements of the HAPs.

**Acknowledgment**

The research was supported by the QCHAPS project, funded through UK Engineering and Physical Sciences Research Council (EPSRC) through project grant references EP/M013472/1 and EP/T001011/1; and the Royal Academy of Engineering through an Early Career Research Fellowship No. RF\201718\1746.

APPENDIX

TABLE II
FIXED-WING HAPs

| Company | HAP name | Aircraft weight | Payload capacity | Payload power | Dimensions | Mobility | Flight duration | Altitude | Availability and timescale |
|---|---|---|---|---|---|---|---|---|---|
| Airbus (UK) | Zephyr S [32] | 75 kg | 2 kg | 50-200 W | 25m wingspan | 55 km/h | 26 days | 21km | Production |
| | Zephyr T [32] | 140 kg | 5 kg | 200-500 W | 33m wingspan | | 45 days | 21km | Development 2016-2019 |
| | Zephyr Future Evolution[32] | | 40 kg | | | | | | Development 2020+ |
| Google (US) | Titan Aerospace (Solara 50) [61] | | 32 kg | | 50m wingspan, 15m length | | | 20km | Abandoned by Google |
| Prismatic (UK) | PHASE-8 [62] | 12 kg | 2 kg | 50 W | 8.75m wingspan | 46 km/h | Days with solar, 8h without | 3km | Production |
| | PHASA-35 [33] | | 15-25 kg | 300-1000 W | | | 1 year for up to 35 °latitude | 16-21 km | Production |
| AlphaLink (GER) | AlphaLink (multi-body) [63] | | 24 kg each (450 kg total) | | 21m wingspan each, can connect up to 10 wings | | 10 days (one year in the future) | 20 km | First model of coupling 3 wings |
| UAVOS (US) | ApusDuo [64] | 23 kg | 2 kg | | 15m wingspan | 92 km/h | 1 year at 35 ° latitude | 12-20 km | Production |
| DLR (GER) | Elektra-2 [35] | 420 kg | 120 kg | 5000 W | 25m wingspan | 70 km/h | Almost unlimited | 20 km | Production |
| Facebook (US) | Aquila [65] | 400 kg | | | 43m wingspan | 128 km/h | 90 days | 18-27 km | Abandoned by Facebook |
| Boeing (US) | Odysseus [34] | | 25 kg | 250 W | 74m wingspan | 160 km/h | Months | 20 km | Test flight in 2019 |
| NASA (US) | Centurion [66] | 560 kg | 272 kg | | 63m wingspan | 33 km/h | 90 minutes test flight | 30 km | Test flight in 1998 |
| | Helios [67] | 600 kg | 330 kg | | 75m wingspan | 43 km/h | 24 hours | 30 km | Destroyed in 2003 |
| | Global Hawk [68] | 11.6 ton maximum | 680 kg | | 35m wingspan, 13.5m length, 4.6m height | 620 km/h | 31 hours with 7 ton fuel | 20 km | Operation |
| Ordnance Survey (UK) | Astigan [69] | 149 kg | 25 kg | | 38m wingspan | | 90 days | 21 km | Low-altitude test 2016, launch 2020 |
| HAPSMobile/SoftBank (JPN+US) | Sunglider [70] | | | | 78m wingspan | 110 km/h | Months | 20 km | Production in 2023 |
| Stratospheric Platforms (UK) | Stratospheric Platforms HAP [36] | 3.5 tons | 140 kg | 20 kW | 60 m wingspan | | 9 days | 20 km | Prototype test flight in 2022 |

* All HAPs are solar powered except the Global Hawk (fuel) and the Stratospheric Platforms HAP (hydrogen). All information is available in the public domain.

TABLE III
LIGHTER-THAN-AIR AERIAL PLATFORMS

| Company | HAP name | HAP type | Aircraft weight | Payload capacity | Payload power | Dimensions | Altitude | Availability and timescale |
|---|---|---|---|---|---|---|---|---|
| Zero 2 Infinity (Spain) | Bloonstar [71] | Rockoon | | 140 kg | | | LEO/SSO | Development |
| | Bloon [72] | Helium balloon | | 6 persons | | | 36km | Operation |
| CNES (FR) | Stratospheric balloon [73] | Helium balloon | 754 kg | 400 kg | | | 37km | Several test flights |
| Google (US) | Loon [23] | Helium balloon | | 10 kg | 100 W with full sun | 74m across, 12m tall | 25 km | Operation |
| Thales Alenia Space (FR) | Stratobus [38] | Airship | 5 ton | 250 kg | 5 kW | 140m length, 32m diameter | 20 km | Down scaled prototype reaching the market in 2020 |
| Avealto, Ltd (UK) | Ascender 28 [74] | Airship | | | | 28m length (60m final) | 25 km | Development |
| Sceye (US) | Sceye [75] | Airship | | | | | 20 km | Down scaled prototype tested in October 2019 |
| Lockheed Martin (US) | 420K Aerostats [76] | Airship | | 1 ton | | 64m length, 12000m$^3$ capacity | 4600 m | Operation |
| | 74K Aerostats [76] | Tethered platform | | 500 kg | | 35m length, 2100m$^3$ capacity | Tethered | Operation |
| Allsopp Helikites (UK) | Helikite [39] | Tethered platform | | Up to 30 kg | | Up to 64 m$^3$ capacity | Up to 1.5 km | Production |

* All platforms are solar powered except the tethered ones. All information in this table is available in the public domain.